\documentclass[12pt]{article}
\usepackage{amscd,amssymb,amsmath,latexsym,enumerate}
\usepackage[mathscr]{euscript}
\usepackage{epsfig}
\usepackage{fancybox}
\usepackage{verbatim}

\usepackage{color}

\title{Quantization of interface currents}

\author{Motoko Kotani$^1$, Hermann Schulz-Baldes$^2$, Carlos Villegas-Blas$^3$
\\
\\
{\small $^1$ AIMR, Tohoku Universtity, Sendai, Japan}
\\
{\small $^2$ Department Mathematik, Universit\"at Erlangen-N\"urnberg, Germany}
\\
{\small $^3$ Instituto de Matematicas, Cuernavaca, UNAM, Mexico}
}

\date{ }

\newtheorem{theo}{Theorem}

\newtheorem{proposi}{Proposition}
\newtheorem{lemma}{Lemma}
\newtheorem{coro}{Corollary}

\newcommand{\CM}{{\mathbb C}}
\newcommand{\NM}{{\mathbb N}}
\newcommand{\RM}{{\mathbb R}}

\newcommand{\ZM}{{\mathbb Z}}

\newcommand{\Ee}{{\cal E}}

\newcommand{\PP}{{\bf P}}

\newcommand{\Dd}{{\cal D}}

\newcommand{\Tr}{\mbox{\rm Tr}}
\newcommand{\Tt}{{\cal T}}

\newcommand{\Hh}{{\cal H}}

\newcommand{\one}{{\bf 1}}

\newcommand{\Ind}{{\rm Ind}} 
\newcommand{\Wind}{{\rm Wind}}

\newcommand{\Che}{{\rm Ch}}

\newcommand{\Ah}{A} 
\newcommand{\Bh}{B} 
\newcommand{\Ch}{C} 
\newcommand{\Uh}{U} 
 
\newcommand{\TVh}{\widehat{\cal T}}

\begin{document}

\maketitle

\begin{abstract}
At the interface of two two-dimensional quantum systems, there may exist interface currents similar to edge currents in quantum Hall systems. It is proved that these interface currents are  macroscopically quantized by an integer that is given by the difference of the Chern numbers of the two systems. It is also argued that at the interface between two time-reversal invariant systems with half-integer spin, one of which is trivial and the other non-trivial, there are dissipationless spin-polarized interface currents.
\end{abstract}


\section{Overview}
\label{sec-overview}

It is well-known that boundaries of quantum systems may lead to surface modes which can carry edge currents. The prime example is a quantum Hall system. Less studied are interfaces between two different materials, but also such interfaces may lead to modes carrying currents. Examples are magnetic walls obtained from different magnetic fields in two half-spaces \cite{Iwa,RP}. On a classical level, the two cyclotron orbits of different radius and/or different orientation lead to the so-called snake orbits which are extended along the interface. Under certain assumptions, it is presumably possible to show by the Mourre commutator method as in \cite{BP,FGW} that the spectrum of these interface modes is absolutely continuous even if a small random potential is added. What is more interesting is to analyze the current density of these modes, in particular, whether a quantization holds as does for quantum Hall systems \cite{SKR,KRS,EG}. In fact, it will be shown below that the interface currents are quantized  (Theorem~\ref{theo-interfacchannels}) and, moreover, that the interface channel number is equal to the difference of the Chern numbers of the two systems (Theorem~\ref{theo-bulkinterface}). All this is exhibited to be robust against random perturbations and to be quite model independent. For example, non-trivial currents flow at the interface between a disordered Haldane model \cite{Hal} and some periodic operator with trivial band topology, even though neither model has a net magnetic field (see example below). Let us point out that the proofs below also cover the case of a half-space if one of the two materials is chosen to be the vacuum.

\vspace{.2cm}

There is a prior rigorous study with roughly the same outcome \cite{DGR}, which, however, only considered  Landau operators and a restricted class of perturbations. Furthermore, the mathematical techniques differ. Here it is first shown that interface channels are well-defined and described by a quasi-one-dimensional index theorem of Noether-Gohberg-Krein type (Theorem~\ref{theo-interfacchannels}). This part of the analysis follows relatively closely a prior paper \cite{KRS}, but several technical points are considerably improved. The second main theorem (Theorem~\ref{theo-bulkinterface}) is then proved by a homotopy argument decoupling the system into two half-plane models for which the bulk-edge correspondence \cite{Hat} as proved in \cite{KRS,EG} can be applied.

\vspace{.2cm}

This set of ideas can also be applied to time-reversal invariant systems with half-integer spin. For such systems the Chern numbers and hence direct currents vanish, but they may nevertheless have non-trivial topology and resulting spin-polarized edge currents which are not susceptible to Anderson localization. The prime example is a quantum spin Hall system described by the Kane-Mele model \cite{KM} which consists of two superposed Haldane models coupled by a Rashba spin-orbit term breaking any other symmetry apart from time-reversal symmetry (see also \cite{ASV} for a detailed description and analysis). The topology is a $\ZM_2$ invariant \cite{KM}  which in an adequate index theoretic formulation \cite{Sch2} is again stable under random perturbations. The surface modes of a periodic system can be studied either numerically \cite{KM} or by  transfer matrix methods \cite{ASV}. For these periodic systems the surface modes carry spin-polarized edge currents which are, however, not quantized as soon as the Rashba coupling is turned on. Even though there is no quantization, it is possible, though, to prove by a controlled perturbation theory that these spin-polarized edge currents vary continuously in the disorder and the Rashba coupling parameter (actually, even in a Zeeman term breaking time-reversal symmetry). This indeed shows that the spin-polarized surface modes do not localize in a weak random potential as one might erroneously expect from the theory of quasi-one-dimensional random systems, provided the $\ZM_2$ invariant is non-trivial. Now let us consider an interface between two time-reversal symmetric models (say a Kane-Mele model and a topologically trivial one). The idea is then to first apply the results of the present paper to a system without Rashba coupling (namely a system which conserves the $s^z$-component of the spin so that both components can be dealt with separately) and then run the perturbative argument from \cite{Sch2}. As a result, the spin-polarized interface currents remain non-vanishing under the circumstances described above. As a full-fledged argument is rather lengthy and repetitive, but not difficult, no further details of proof are given and even the precise formulation of the result is left to the interested expert.

\vspace{.2cm}

Let us now continue with a precise formulation of the results of the present paper. Given is a compact space of disorder configurations $\Omega$ furnished with a $\ZM^2$-action $T=(T_1,T_2):\ZM^2\times\Omega\to\Omega$ and an invariant and ergodic probability measure $\PP$. Then let $H_\pm=(H_{\pm,\omega})_{\omega\in\Omega}$ be two covariant families of Hamiltonians on $\ell^2(\ZM^2)$, namely  
\begin{equation}
\label{eq-covrel}
(S^{B_\pm}_{j})^* \,H_{\pm,\omega}\,S^{B_\pm}_{j}
\;=\;
H_{\pm,T_j\omega}
\;,
\qquad
j=1,2
\;.
\end{equation}
Here $S^{B_\pm}_{1}$ and $S^{B_\pm}_{2}$ are the magnetic translations on $\ell^2(\ZM^2)$ with  two possibly different constant magnetic fields $B_+,B_-\in\RM$ ({\it e.g.} \cite{BES}). These operators satisfy
$$
S^{B_\pm}_{1}S^{B_\pm}_{2}
\;=\;
e^{\imath B_\pm}\,S^{B_\pm}_{2}S^{B_\pm}_{1}
\;,
$$
and the choice of gauge is not relevant here. It is possible to allow finite dimensional fibers over every site of $\ZM^2$, but this will not be done here for sake of simplicity (it is necessary, however, in order to deal with the Haldane and Kane-Mele model). It will be assumed that $H_\pm$ are short ranged, say that the matrix elements of $\langle n|H_\pm|m\rangle$ vanish for $n,m>R$ for some $R>0$. Due to covariance, the spectra $\sigma(H_{\pm,\omega})$ are $\PP$-almost surely equal to sets denoted by $\sigma(H_\pm)$. All of the below will be done under the following standing assumption.

\vspace{.2cm}

\noindent {\bf Gap condition:} {\sl The Fermi level $E_F$ lies in an open interval $\Delta\subset\RM$ having an empty intersection with the} ({\sl almost sure}) {\sl spectra} $\sigma(H_\pm)$.

\vspace{.2cm}

The aim in the following is to restrict the two Hamiltonians $H_\pm$ to the upper and lower half-space respectively and then to couple the two restrictions. Hence let $\Pi_{2,\pm}$ denote the partial isometries from $\ell^2(\ZM^2)$ onto the subspace $\Hh_\pm=\ell^2(\ZM\times\ZM_\pm)$ where $\ZM_+=\ZM\cap[0,\infty)$ and $\ZM_-=\ZM\cap(-\infty,0)$. One then has $\Hh=\Hh_+\oplus\Hh_-$ and the half-space restrictions of the Hamiltonians are
$$
\widehat{H}_{\pm,\omega}
\;=\;
\Pi_{2,\pm}\,H_{\pm,\omega}\,(\Pi_{2,\pm})^*
\;,
$$
seen as operators on $\Hh_\pm$ respectively. Furthermore let $K=(K_{\omega})_{\omega\in\Omega}$ be a family of operators on $\ell^2(\ZM^2)$ which are supported on a strip $\ZM\times [-N,N]$ of width $N$, namely $\langle n|K_\omega|m\rangle=0$ if either $n\not\in\ZM\times [-N,N]$ or $m\not\in\ZM\times [-N,N]$, which is moreover covariant in the $1$-direction.  From all these data is constructed a family $H=(H_\omega)_{\omega\in\Omega}$ of Hamiltonians on $\ell^2(\ZM^2)$ by
\begin{equation}
\label{eq-Hamform}
{H}_{\omega}
\;=\;
\widehat{H}_{+,\omega}
\oplus
\widehat{H}_{-,\omega}
\;+\;
K_{\omega}
\;.
\end{equation}
These Hamiltonians model an interface between two materials submitted to different constant magnetic fields. Let us note that they are only covariant in the $1$-direction which is parallel to the interface. It is important to realize that the Gap hypothesis does {\it not} imply that $\Delta$ has an empty intersection with the spectrum of $H_\omega$. In fact, typically the full interval $\Delta$ is contained in $\sigma(H_\omega)$. However, the associated states are interface states in the sense of the following 

\begin{proposi}
\label{prop-decay} 
Let $g$ be a smooth function supported in $\Delta$. Then for all $\alpha>0$ there exists a constant $C$ such that for any $\omega\in\Omega$, $n=(n_1,n_2)\in\ZM^2$ and $m=(m_1,m_2)\in\ZM^2$,
\begin{equation}
\label{eq-interfacedecay}
|\langle n|g(H_\omega)|m\rangle|
\;\leq\;
\frac{C}{1+|n_2|^\alpha+|m_2|^\alpha+|n_1-m_1|^\alpha}
\;.
\end{equation}
\end{proposi}

\vspace{.2cm}

This implies that $g(H)=(g(H_\omega))_{\omega\in\Omega}$ is $\widehat{\Tt}$-traceclass, where, for a family $A=(A_\omega)_{\omega\in\Omega}$ of operators that are covariant in the $1$-direction, the trace $\widehat{\Tt}$ is defined by
\begin{equation}
\label{eq-that}
\widehat{\Tt}(A)
\;=\;
\int \PP(d\omega)\;\sum_{n_2\in\ZM}\,\langle 0,n_2|A_\omega|0,n_2\rangle
\;,
\end{equation}
strictly speaking on a dense ideal in the algebra $\Ee$ introduced below. The current operator in the $1$-direction is now defined in terms of the position operator $X_1$ by $J_{1,\omega}=\imath[X_1,H_\omega]$. For the finite range hopping Hamiltonian, this operator is bounded. Hence by the above, the interface current density $\widehat{\Tt}(g(H)J_{1})$ is well-defined. 

\begin{theo} 
\label{theo-interfacchannels} 
Suppose that the Gap hypothesis holds. Let $g$ be a positive smooth function supported in $\Delta$ which is of unit integral and set $G(E)=\int^E_{-\infty}dE'\,g(E')$ and introduce $U=(U_\omega)_{\omega\in\Omega}$ by
$$
U_\omega
\;=\;
\exp(2\pi\imath\,G(H_\omega))
\;.
$$
With the surjective partial isometry $\Pi_{1}:\ell^2(\ZM^2)\to\ell^2(\ZM_+\times\ZM)$, the operators $\Pi_{1}U_\omega(\Pi_{1})^*$ on $\ell^2(\ZM_+\times\ZM)$ are Fredholm and their index is almost surely constant in $\omega$ and this almost sure index is given by the non-commutative winding number of $U$:
\begin{equation}
\label{eq-IndWind}
\Ind(\Pi_{1}U_\omega(\Pi_{1})^*)
\;=\;
\,\imath\,
\widehat{\Tt}
\bigl((U^*-\one)\imath[X_1,U]\bigr)
\;.
\end{equation}
Furthermore, the almost sure index is linked to the interface current density by
\begin{equation}
\label{eq-Indinterfacecurr}
2\pi\,\widehat{\Tt}(g(H)J_{1})
\;=\;
\Ind(\Pi_{1}U_\omega(\Pi_{1})^*)
\;.
\end{equation}
\end{theo}

\vspace{.2cm}

The integer appearing in Theorem~\ref{theo-interfacchannels} can be called the interface channel number, in analogy with the edge channel number. The next issue is to calculate this integer from the topology of the bulk states encoded in the Chern numbers of the covariant families $H_\pm$ of Hamiltonians. For $E_F\in\Delta$, let $P_\pm=\chi(H_\pm\leq E_F)$ denote the two Fermi projections. Then their Chern numbers are given by \cite{BES}
$$
\Che(P_\pm)
\;=\;
2\pi\imath\;\int \PP(d\omega)\;\langle 0|P_{\pm,\omega}[[X_1,P_{\pm,\omega}],[X_2,P_{\pm,\omega}]]|0\rangle
\;.
$$

\begin{theo} 
\label{theo-bulkinterface} 
Suppose the gap hypothesis holds. Let $g$ and $G$ be as above.Then
$$
2\pi\,\widehat{\Tt}(g(H)J_{1})
\;=\;
\Che(P_+)
\,-\,
\Che(P_-)
\;.
$$
\end{theo}

Let us provide some non-trivial concrete examples where this theorem applies.

\vspace{.2cm}

\noindent {\bf Example 1} Let  $\widetilde{S}^{B_\pm}_1$ and $\widetilde{S}^{B_\pm}_2$ be the magnetic translations that are dual to $S^{B_\pm}_1$ and $S^{B_\pm}_2$ introduce the notation $(S^{B_\pm})^m=(S^{B_\pm}_1)^{m_1}(S^{B_\pm}_2)^{m_2}$. The dual magnetic translations satisfy the covariance relation \eqref{eq-covrel} and therefore so do the Hamiltonians
$$
H_{\pm,\omega'}
\;=\;
\sum_{m\in\ZM^2} t_{\pm,m} \,(\widetilde{S}^{B_\pm})^m
\;+\;
\lambda\,\sum_{n\in\ZM^2}\,v_n\,|n\rangle\langle n|
\;,
$$
where $t_{\pm,m}\in\RM$ are the hopping amplitudes, non-vanishing only for a finite number of $m$, and the $v_n\in[-1,1]$ are i.i.d. centered random variables which form the random variable $\omega=(v_n)_{n\in\ZM^2}$. Choosing $B_+\not=B_-$ and $E_F$ such that the Gap hypthesis holds and $\Che(P_-)\not=\Che(P_+)$ is possible (as is easily realized by looking at the Hofstadter butterfly). The coupling $K_\omega$ in  \eqref{eq-Hamform} can {\it e.g.} be chosen non-random. Theorem~\ref{theo-bulkinterface} now guarantees non-vanishing quantized interface currents.
\hfill $\diamond$

\vspace{.2cm}

\noindent {\bf Example 2}  Let $H_+$ be the Haldane model \cite{Hal}  and $H_-$ the discrete Laplacian on the honeycomb lattice with a staggered potential opening a gap at the Fermi energy $E_F=0$. Then one may add a small random potential and some coupling $K_\omega$ as above. As there are different Chern numbers $\Che(P_+)=1$ and $\Che(P_-)=0$, there are again non-vanishing interface currents even though both $H_+$ and $H_-$ have a vanishing net magnetic field.
\hfill $\diamond$

\vspace{.3cm}

\noindent {\bf Acknowledgements:} We are thankful for financial support from AIMR, DFG and project PAPIIT-UNAM IN 106812. 

\section{Proof of Proposition~\ref{prop-decay}}

Let us begin by recalling the Helffer-Sj\"ostrand functional calculus for a compactly supported  smooth function $g:\RM\to\CM$ of an arbitrary self-adjoint and bounded operator $H$:
\begin{equation}
\label{eq-HS}
g(H)
\;=\;
\frac{-1}{2\pi}\;
\int_{\RM^2}
dx\,dy\;
\partial_{\overline{z}}\widetilde{g}(x,y)\;(z-H)^{-1}\;,
\qquad
z=x+\imath y
\;,
\end{equation}
where, for some $N\geq 1$,
$$
\widetilde{g}(x,y)
\;=\;
\sum_{n=0}^N \;g^{(n)}(x)\;\frac{(\imath y)^n}{n!}\;\chi(y)\;,
$$
with some smooth, even, compactly supported function $\chi:(-1,1)\to [0,1]$ which is equal to $1$ on $[-\delta,\delta]$. One can always choose $g$ constant (say vanishing) outside of the spectrum of $H$. The integral in \eqref{eq-HS} is norm convergent and the function $\widetilde{g}$, called a quasi-analytic extension of $g$, satisfies
$$
\partial_{\overline{z}}\widetilde{g}(x,y)
\; = \;
g^{(N+1)}(x)\;\frac{(\imath y)^N}{N!}\;\chi(y)
\,+\,\imath\,
\sum_{n=0}^N g^{(n)}(x)\;\frac{(\imath y)^n}{n!}\;\chi'(y)
\;,
$$
so that, in particular, uniformly in $x,y$, 
\begin{equation}
\label{eq-gtildebound}
|\partial_{\overline{z}}\widetilde{g}(x,y)|\;\leq \;C_0\,\|g\|_{N+1}\,|y|^{N}
\;,
\end{equation}
where $\|g\|_{N+1}$ denotes the usual norm on $(N+1)$-times differentiable functions.

\vspace{.2cm}

\noindent {\bf Proof } of Proposition~\ref{prop-decay}. The estimate is pointwise in $\omega$, so let us drop the indices $\omega,\omega',\omega''$ on the Hamiltonian which is of the form $H=\widehat{H}_+\oplus\widehat{H}_-+K$ as in \eqref{eq-Hamform}. The operator $g(H)$ will be written with the Helffer-Sjorstrand formula. The resolvent identity for $H$ shows
$$
\frac{1}{z-H}
\;=\;
\frac{1}{z-\widehat{H}_+}\oplus\frac{1}{z+\widehat{H}_-}
\;+\;
\frac{1}{z-\widehat{H}_+\oplus\widehat{H}_-}
\,K\,
\frac{1}{z-H}
\;,
$$
where the direct sum is w.r.t. the decomposition $\Hh=\Hh_+\oplus\Hh_-$. Replaced in \eqref{eq-HS}, one obtains
\begin{equation}
\label{eq-gdecomp}
g(H)
\;=\;
g(\widehat{H}_+)\oplus g(\widehat{H}_-)
\,-\,
\frac{1}{2\pi}\;
\int_{\RM^2}
dx\,dy\;
\partial_{\overline{z}}\widetilde{g}(x,y)\;
\frac{1}{z-\widehat{H}_+\oplus\widehat{H}_-}
\,K\,
\frac{1}{z-H}
\;.
\end{equation}
By hypothesis, $g$ is supported in a gap of both $H_+$ and $H_-$. Therefore $g(\widehat{H}_+)$ and $g(\widehat{H}_-)$ both satisfy the estimate \eqref{eq-interfacedecay} by the results of \cite{EG,Sch} (which again follows from Helffer-Sj\"orstrand, the geometric resolvent identity combined with the Combes-Thomas estimate as below). Hence it only remains to deal with the matrix elements of the second summand in \eqref{eq-gdecomp}. For that let us recall the Combes-Thomas estimate which states that there is an $\eta>0$ and $C$ such that
$$
|\langle k|(z-H)^{-1}|l\rangle|
\;\leq\;
\frac{C}{|y|}\;e^{-\eta|y| |k-l|}
\;,
\qquad
k,l\in\ZM^2
\;.
$$
A detailed proof in the present context can be found in \cite[Proposition~2]{DDS}. The same estimate holds for the resolvent of $\widehat{H}_+\oplus\widehat{H}_-$ (only a decay of the hopping elements is needed).
$$
|\langle n|g(H)-
g(\widehat{H}_+)\oplus g(\widehat{H}_-)|m\rangle|
\,\leq\,
\int_{\RM^2}
dx\,dy\;
|\partial_{\overline{z}}\widetilde{g}(x,y)|\,
\frac{C^2}{y^2}
\;
\sum_{k,l\in\ZM^2}\,
e^{-\eta|y| |n-k|}\,
|\langle k|K|l\rangle|
\,e^{-\eta|y| |l-m|}
\;.
$$
Now the operator $K$  has non-vanishing matrix elements only in a finite distance away from the boundary (thus $k_2$ and $l_2$ are close to $0$) and is finite range in $k_1-l_1$. Thus invoking the bound \eqref{eq-gtildebound} with $N=\alpha+2$, one deduces with some care that this term also satisfies the bound \eqref{eq-interfacedecay}. 
\hfill $\Box$

\section{Proof of Theorem~\ref{theo-interfacchannels}}

\subsection{Definition of the edge algebra}

Let $\Ee_0$ be the set of families $A=(A_\omega)_{\omega\in\Omega}$ of operators on $\ell^2(\ZM^2)$ which are covariant in the $1$-direction and of finite support in the $2$-direction and finite hopping distance in the $1$-direction, namely 
\begin{equation}
\label{eq-covariance}
\langle n|A_\omega|m\rangle
\;=\;
\langle n-k|A_{T^k\omega}|m-k\rangle
\;,
\qquad
n,m\in\ZM^2\,,\;k\in \ZM\times\{0\}\;,
\end{equation}
and satisfy for some constant $C$
$$
\langle n|A_\omega|m\rangle\;=\;0
\qquad
|n_2|\geq C\;\;\mbox{or}\;\;
|m_2|\geq C\;\;\mbox{or}\;\;
|n_1-m_1|\geq C\;.
$$
Then $\Ee_0$ is a $*$-algebra and its C$^*$-closure $\Ee$ is called the edge algebra. On $\Ee_0$ is defined the derivative
$$
(\nabla_1A)_\omega
\;=\;
\imath [X_1,A_\omega]
\;.
$$
Furthermore, one has a trace on $\Ee_0$ defined by \eqref{eq-that}. Both $\nabla_1$ and $\TVh$ extend to dense subsets of $\Ee$ (the respective domains). The set of operators for which $(\nabla_1)^kA\in\Ee$ is denoted by $C^k(\Ee)$, and the $\TVh$-traceclass operators by $L^1(\Ee,\TVh)$.

\subsection{1-cocycles over the edge algebra}
\label{subsec-cocycle}

The following expression is well-defined and finite for
$\Ah,\Bh\in\Ee_0$:

\begin{equation}
\label{eq-xi1}
\xi_1(\Ah,\Bh)
\;=\;
\imath\;\TVh(\Ah\nabla_1\Bh)
\mbox{ . }
\end{equation}

\noindent Actually $\xi_1$ can be defined on a much wider class of operators. For the purpose of this work, it is sufficient to consider elements in $\Dd= C^{2}(\Ee)\cap L^1(\Ee,\TVh)$. Note that the product rule for $\nabla_1$ and the ideal property of the $\TVh$-trace-class operators imply that $\Dd$ is a $*$-algebra. It becomes a normed algebra when endowed with the norm:

$$
\|\Ah\|_\Dd
\;=\;
\|\Ah\| + 2 \|\nabla_1 \Ah\|
+\|\nabla_1^2 \Ah\|
+\TVh(|\Ah|)
\mbox{ . }
$$

\vspace{.2cm}

\begin{lemma}
\label{lem-cocycle}
$\xi_1$ is a cyclic 1-cocycle on $\Dd$,
notably it is cyclic and closed under the
Hochschild boundary operator $b$ defined by $b\xi_1(\Ah,\Bh,\Ch)
=\xi_1(\Ah\Bh,\Ch)-\xi_1(\Ah,\Bh\Ch)+\xi_1(\Ch \Ah,\Bh)$:

\vspace{.1cm}

\noindent {\rm (i)} $\xi_1(\Ah,\Bh)=-\xi_1(\Bh,\Ah)$ for all
$\Ah,\Bh\in\Dd$.

\vspace{.1cm}

\noindent {\rm (ii)} $0=b\xi_1(\Ah,\Bh,\Ch)$
for all $\Ah,\Bh,\Ch\in\Dd$.

\end{lemma}

\vspace{.2cm}

\noindent{\bf Proof.} Both algebraic identities can be verified using the product rule for the derivation $\nabla_1$ and the invariance of the trace $\TVh$ under $\nabla_1$.
\hfill $\Box$

\vspace{.2cm}

Next let us introduce another 1-cocycle on the unitalization $\widetilde{\Dd}=\Dd\cup \CM\,\one$ by setting
$$
\zeta_1(\Ah,\Bh)
\;=\;
\int \PP(d\omega)\;\zeta^\omega_1(\Ah,\Bh)
\mbox{ , }
\qquad
\Ah,\Bh\in\widetilde{\Dd}
\mbox{ , }
$$
\noindent where
\begin{equation}
\label{eq-zetacyc}
\zeta^\omega_1(\Ah,\Bh)
\;=\;
\frac{1}{4}\;
\Tr_{\ell^2(\ZM\times\NM)}
\left(
\frac{X_1}{|X_1|}
\left[\frac{X_1}{|X_1|},\Ah_\omega\right]
\left[\frac{X_1}{|X_1|},\Bh_\omega\right]
\right)
\mbox{ ,}
\end{equation}
with the convention $\frac{X_1}{|X_1|}|0,n_2\rangle=|0,n_2\rangle$.

\begin{proposi}
\label{prop-cocyclelink1}
On $\Dd$, we have $\zeta_1=\xi_1$.
\end{proposi}

\noindent{\bf Proof.}
It is sufficient to show the equality for the dense subalgebra $\Ee_0\subset\Dd$ because both $\zeta_1$ and $\xi_1$ are continuous with respect to $\|\,.\,\|_\Dd$. A direct calculation shows that for $\Ah,\Bh\in\Ee_0$:
$$
\zeta_1(\Ah,\Bh)
\;=\;
-\frac{1}{4}
\int \PP(d\omega)\;
\sum_{m\in \ZM\times \NM}
\sum_{l\in \ZM\times \NM}
\mbox{sgn}(m_1)(\mbox{sgn}(m_1)-\mbox{sgn}(l_1))^2
\langle m|\Ah_\omega|l\rangle\,
\langle l|\Bh_\omega| m\rangle
\mbox{ . }
$$
\noindent Because $\Ah\in \Ee_0$, the sum over $m_1\in \ZM$ actually only contains a finite number of non-zero elements, and can thus be exchanged with the integral over $\PP$. Then we make the change of variables $n_1=l_1-m_1$ and use the covariance relation in order to obtain:
\begin{eqnarray}
\zeta_1(\Ah,\Bh)
& = &
-\frac{1}{4}
\sum_{m_1\in\ZM}\int \PP(d\omega)\;
\sum_{m_2,l_2\in \NM}
\sum_{n_1\in \ZM}
\mbox{sgn}(m_1)\;(\mbox{sgn}(m_1)-\mbox{sgn}(m_1+n_1))^2\,\cdot
\nonumber
\\
& &
\;\;\;\;\;\;\;\;\;\;\;\;\;\;\;\;\;\;\;\;\cdot\;
\langle 0,m_2|\Ah_{T^{(-m_1,0)}\omega}|n_1,l_2\rangle\,
\langle n_1,l_2 |\Bh_{T^{(-m_1,0)}\omega}|0,m_2\rangle
\mbox{ . }
\nonumber
\end{eqnarray}
\noindent Next, by invariance of the measure $\PP$, we can replace
$T^{(-m_1,0)}\omega$ by $\omega$. Then we change the sum over $m_1$ and
the integral over $\PP$ again and use the identity
$$
\sum_{m_1\in\ZM}
\mbox{sgn}(m_1)\;(\mbox{sgn}(m_1)-\mbox{sgn}(m_1+n_1))^2
\;=\;
-\,4\,n_1
\mbox{ . }
$$
\noindent By definition of $\nabla_1$, one therefore has
$$
\zeta_1(\Ah,\Bh)
\;=\;
\imath\,\int \PP(d\omega)\;
\sum_{m_2\in \NM}
\langle 0,m_2|\Ah_\omega
(\nabla_1\Bh)_\omega|0,m_2\rangle
\mbox{ , }
$$
\noindent which is precisely $\xi_1(\Ah,\Bh)$.
\hfill $\Box$

\vspace{.2cm}

Finally let us introduce a further 1-cocyle on $\widetilde{\Dd}$
using the surjective partial isometry $\Pi_1:\ell^2(\ZM^2)\to\ell^2(\ZM_+\times\ZM)$:
$$
\eta_1(\Ah,\Bh)
\;=\;
\int \PP(d\omega)\;\eta^\omega_1(\Ah,\Bh)
\mbox{ , }
\qquad
\Ah,\Bh\in\tilde{\Dd}
\mbox{ , }
$$
\noindent where
\begin{eqnarray}
\label{eq-cocycle3}
\eta^\omega_1(\Ah,\Bh)
& = &
\Tr_{\ell^2(\ZM_+\times\ZM)}
\big(\Pi_{1}\Bh_\omega\Ah_\omega(\Pi_{1})^*-
\Pi_{1}\Bh_\omega(\Pi_{1})^*\Pi_1\Ah_\omega(\Pi_{1})^*\big)
\\
&  &
-\Tr_{\ell^2(\ZM_+\times\ZM)}
\big(\Pi_{1}\Ah_\omega\Bh_\omega(\Pi_{1})^*-
\Pi_{1}\Ah_\omega(\Pi_{1})^*\Pi_1\Bh_\omega(\Pi_{1})^*\big)
\nonumber
\mbox{ . }
\end{eqnarray}

\begin{proposi}
\label{prop-cocyclelink2}
On $\widetilde{\Dd}$, both expressions in {\rm (\ref{eq-cocycle3})}
are finite and one has $\eta^\omega_1=\zeta^\omega_1$ for all
$\omega\in\Omega$.
\end{proposi}

\vspace{.2cm}

\noindent{\bf Proof.} Some algebra shows
\begin{equation}
\label{algiden}
\Pi_{1}\Ah_\omega\Bh_\omega(\Pi_{1})^*-
\Pi_{1}\Ah_\omega(\Pi_{1})^*\Pi_1\Bh_\omega(\Pi_{1})^*
\;=\;
-\frac{1}{4}
\Pi_{1}
\left[\frac{X_1}{|X_1|},\Ah_\omega\right]
\left[\frac{X_1}{|X_1|},\Bh_\omega\right]
(\Pi_1)^*
\mbox{ . }
\end{equation}
\noindent Note that if $\left[\frac{X_1}{|X_1|},\Ah_\omega\right]
\left[\frac{X_1}{|X_1|},\Bh_\omega\right]$ is trace-class, so is the
left-hand side (this is the case for $\Ah,\Bh\in\widetilde{\Dd}$).
Hence using the same identity with $\Ah$ and $\Bh$
exchanged, we obtain:
\begin{eqnarray}
\eta^\omega_1
(\Ah,\Bh)
\! & \! = \! & \!
\frac{1}{4}\,
\Tr_{\ell^2(\ZM_+\times\ZM)}
\left(
\Pi_1
\left[\frac{X_1}{|X_1|},\Ah_\omega\right]
\left[\frac{X_1}{|X_1|},\Bh_\omega\right]
(\Pi_1)^*-
\Pi_1
\left[\frac{X_1}{|X_1|},\Bh_\omega\right]
\left[\frac{X_1}{|X_1|},\Ah_\omega\right]
(\Pi_1)^*\right)
\nonumber
\\
\! & \! = \! & \!
\frac{1}{8}\,
\Tr_{\ell^2(\ZM\times\ZM)}
\left(
\frac{X_1}{|X_1|}
\left[\frac{X_1}{|X_1|},\Ah_\omega\right]
\left[\frac{X_1}{|X_1|},\Bh_\omega\right]
-
\frac{X_1}{|X_1|}
\left[\frac{X_1}{|X_1|},\Bh_\omega\right]
\left[\frac{X_1}{|X_1|},\Ah_\omega\right]
\right).
\nonumber
\end{eqnarray}
\noindent Note here that the second equality holds because both $\left[\frac{X_1}{|X_1|},\Ah_\omega\right]$ and $\left[\frac{X_1}{|X_1|},\Bh_\omega\right]$ are Hilbert-Schmidt operators. From the above we deduce that
$$
\eta^\omega_1
(\Ah,\Bh)
\;=\;
\frac{1}{2} \;(\zeta^\omega_1
(\Ah,\Bh)-\zeta^\omega_1(\Bh,\Ah))
\mbox{ , }
$$
\noindent and the cyclicity property of $\zeta^\omega_1$ allows to conclude.
\hfill $\Box$

\vspace{.2cm}

\begin{coro}
\label{coro-cocyclelink}
On $\Dd$, we have $\xi_1=\zeta_1=\eta_1$.
\end{coro}

\subsection{Index theorems}
\label{subsec-index}

\begin{proposi}
\label{prop-oddindex1} Suppose $($only for the purpose of this
proposition$)$ that $\PP$ is ergodic w.r.t. the $\ZM$-action
$T_1$. Let $\Ah\in\widetilde{\Dd}$ be unitary. Then $\Pi_{1}
\Ah_\omega(\Pi_{1})^*$ is $\PP$-almost surely a Fredholm
operator on $\ell^2(\ZM_+\times\ZM)$ the index of which is
$\PP$-almost surely independent of $\omega\in\Omega$. Its common
value is equal to $\xi_1(\Ah-\alpha, \Ah^{*}-\overline{\alpha})$
whenever $\Ah-\alpha\in\Dd$, $\alpha\in\CM$, $\alpha\neq 0$.
\end{proposi}

\vspace{.2cm}

\noindent{\bf Proof.} Because $\Ah\in\widetilde{\Dd}$, Proposition
\ref{prop-cocyclelink2} implies that $\eta^\omega_1(\Ah,\Ah^{*})<\infty$
for $\PP$-almost all $\omega\in\Omega$ and that $\one-\Pi_{1}\Ah_\omega^*(\Pi_{1})^*\Pi_{1}\Ah_\omega(\Pi_{1})^*$ and $\one-\Pi_{1}\Ah_\omega(\Pi_{1})^*\Pi_{1}\Ah_\omega^*(\Pi_{1})^*$ are both traceclass. Hence $\Pi_{1}\Ah_\omega(\Pi_{1})^*$ is a Fredholm
operator on $\Pi_{1}\ell^2(\ZM^2) = \ell^2(\ZM_+\times\ZM)$ and
by the well-known Calderon-Fedosov formula its index is equal to $\eta^\omega_1(\Ah,\Ah^{*})$, see {\it e.g.} the appendix to \cite{Con}. Because
$$
\Pi_{1}\Ah_{T^{(a,0)}\omega}
(\Pi_{1})^*\mid_{\ell^2(\ZM_+\times\ZM)}
\;=\;
\Pi_{1}\Ah_\omega(\Pi_{1})^* + K
\mid_{\Uh(a,0)\ell^2(\ZM_+\times\ZM)=\ell^2(\ZM_+\times\ZM) }
\mbox{ , }
$$
\noindent where $K$ is a compact operator on
$\ell^2(\ZM_+\times\ZM)$ and the Fredholm index is invariant under
compact perturbations, we see that the index is $T_1$-translation
invariant in $\omega\in\Omega$. Hence it is $\PP$-almost surely
constant by the ergodicity of $\PP$ with respect to $T_1$.  As
$\eta^\omega_1(\Ah,\Ah^{*})=
\eta^\omega_1(\Ah-\alpha,\Ah^{*}-\overline{\alpha})$, Corollary
\ref{coro-cocyclelink} implies that the almost sure index is equal
to $\xi_1(\Ah-\alpha, \Ah^{*}-\overline{\alpha})$. \hfill $\Box$

\vspace{.2cm}

In our context, the measure $\PP$ is only ergodic w.r.t. the
$\ZM^2$-action $T=(T_1,T_2)$. However, this is sufficient to give an almost
sure index for certain elements in $\widetilde{\Dd}$, notably those in the
image of the exponential map.

\vspace{.2cm}

\begin{proposi}
\label{prop-oddindex2}
Let $H=(H_\omega)_{\omega\in\Omega}$ be of the form {\rm \eqref{eq-Hamform}} and $G$ be a real $C^4$ function
with values in $[0,1]$, equal to $0$ or $1$ outside of $\Delta$.
Set ${U_\omega}=\exp(-2\pi\imath \,G({H_\omega}))$.
Then $\Pi_{1}U_\omega(\Pi_{1})^*$ is $\PP$-almost surely a Fredholm operator on
$\ell^2(\ZM_+\times\ZM)$ the index of which is $\PP$-almost surely
independent of $\omega\in\Omega$. The almost sure value is equal to
$\xi_1(U-1, U^{-1}-1)$.
\end{proposi}

\vspace{.2cm}

\noindent{\bf Proof.} By Proposition~\ref{prop-decay},
$U\in\widetilde{\Dd}$. Thus from the proof of Proposition
\ref{prop-oddindex1} follows that $\Pi_{1}
U_\omega(\Pi_{1})^*$ is $\PP$-almost surely
a Fredholm operator and that its index
is $T_1$-invariant. To conclude, we have to show its
$T_2$-invariance.

\vspace{.2cm}

Let us consider $R_\omega=H_{T_2\omega}-H_\omega$. It is clear from \eqref{eq-Hamform} that $R=(R_\omega)_{\omega\in\Omega}\in\Ee_0$. In particular, $R_\omega$ is (part of) a connecting operator allowed in \eqref{eq-Hamform}. For $\lambda\in[0,1]$ let us set
$$
U_\omega(\lambda)
\;=\;
\exp(-2\pi\imath\;G(
H_\omega+\lambda R_\omega))
\mbox{ . }
$$
\noindent This is norm-continuous family (in $\lambda$) for which by the above $\Pi_{1}
U_\omega(\lambda)(\Pi_{1})^*$ is a Fredholm operator. Therefore the index does not change with $\lambda$. As $U_\omega(0)=U_\omega$ and $U_\omega(1)=U_{T_2\omega}$ the proof is concluded.
\hfill $\Box$

\vspace{.2cm}

\noindent {\bf Proof} of identity \eqref{eq-IndWind} of Theorem~\ref{theo-interfacchannels}. This follows immediately from Proposition~\ref{prop-oddindex2}.
\hfill $\Box$

\subsection{Calculation of the edge current}
\label{subsec-currentcalc}

\noindent{\bf Proof} of identity \eqref{eq-Indinterfacecurr} of Theorem~\ref{theo-interfacchannels}. According to the Proposition~\ref{prop-oddindex1},
$$
\Wind\;=\;\imath\;\widehat{\Tt}\bigl( ({U}^*-\one)\nabla_1 {U}\bigr)
\;,
$$
where $\Wind$ denotes the almost sure value of the index. Let us express ${U}$ as an exponential series and use the Leibniz rule:
$$
\widehat{\Tt}(({U}^*-1)\,{\nabla}_1{U}) 
\;=\; 
\imath\;
\sum_{m=1}^\infty
\frac{(2\pi\imath)^m}{m!}
\;\sum_{l=0}^{m-1}\;
\widehat{\Tt}
\left(({U}^*-\one)\,G({{H}})^l\,
{\nabla}_1 G({{H}})\,G({{H}})^{m-l-1}
\right)
\mbox{ , }
$$
\noindent where the trace and the infinite sum could be exchanged
because of the traceclass properties of ${U}-\one$. Due to cyclicity and the fact that 
$[{U},G({{H}})]=0$, each summand is now equal to 
$\widehat{\Tt}(({U}^*-1)\,G({{H}})^{m-1}\,
{\nabla}_1G({{H}}))$. Exchanging again sum and trace and
summing the exponential up again, one gets
$$
\Wind
\;=\;
-2\pi\;\widehat{\Tt}
\left((\one-{U})\,{\nabla}_1G({{H}})\right)
\mbox{ . }
$$
Now let us invoke Lemma~\ref{lem-indexadditivity} below and repeat the same argument for ${U}^k=\exp(-2\pi
\imath \,k\,G({{H}}))$ for  $k\neq 0$,
$$
\Wind\;=\;
\frac{\imath}{k}\;
\widehat{\Tt}\bigl( ({U}^k-\one)^*\nabla_1 {U}^k\bigr)
\;=\;-\,2\pi\;
\widehat{\Tt}
\left((\one-{U}^k)\,{\nabla}_1G({{H}})\right)
\mbox{ . }
$$
\noindent  Writing  $G(E)=\int dt\,\tilde{G}(t)\,e^{-E(1+\imath t)}$ with adequate $\tilde{G}$, the DuHamel formula gives
$$
\Wind
\; = \;
2\pi\;
\int dt\,\tilde{G}(t)\,(1+\imath t)\;
\int^1_0dq\;
\widehat{\Tt}
\left(({U}^k-\one)\,
e^{-(1-q)(1+\imath t){{H}}}
({\nabla}_1 {H})
e^{-q(1+\imath t){{H}}}
\right)
\mbox{ . }
$$
\noindent One therefore finds using
$G'(E)=-\int dt\,(1+\imath t) \,\tilde{G}(t)\,e^{-E(1+\imath t)}$,
for $k\neq 0$,
$$
\Wind
\; = \;
2\pi\;
\widehat{\Tt}
\left(({U}^k-\one)\,G'({{H}})\,
{\nabla}_1{H}
\right)
\mbox{ . }
$$
\noindent For $k=0$, the r.h.s. vanishes.

\vspace{.2cm}

To conclude, let $\phi:[0,1]\to \RM$ be a differentiable function
vanishing at the boundary points $0$ and $1$.
Let its Fourier coefficients be denoted by 
$a_k=\int^1_0 dx \,e^{-2\pi\imath k x}\phi(x)$. Then 
$\sum_k a_k e^{2\pi\imath k x}=\phi(x)$ and, in particular, $\sum_k a_k=0$.
Hence
\begin{eqnarray*}
a_0\;\Wind & = &
-\;\sum_{k\neq 0} a_k\;\Wind
\\
& = &
2\pi\;
\sum_{k} a_k\;
\widehat{\Tt}
\left((\one-{U}^k)\,
G'({{H}})\,{\nabla}_1{H}
\right)
\\
& = &
2\pi\;\widehat{\Tt}
\bigl(\phi(G({H}))\,G'({{H}})\,{\nabla}_1{H}\bigr)
\mbox{ . }
\end{eqnarray*}
\noindent Let now $\phi$ converge to the indicator function of
$[0,1]$. Then $a_0\to 1$, 
while $\phi(G({H}))G'({{H}})\to 
G'({{H}})$ on the other side (the Gibbs phenomenon is
damped). As $J_1=\nabla_1 {H}$, this concludes the proof.
\hfill $\Box$

\begin{lemma}
\label{lem-indexadditivity}
Let ${U}$ and ${V}$ be two unitaries with
${U}-\one\in\Dd$ and ${V}-\one\in\Dd$ so that 
$\Wind({U})=\xi_1({U}-\one, {U}^*-\one)$ and 
$\Wind({V})=\xi_1({V}-\one, {V}^*-\one)$ are well-defined.
Then the unitary ${U}{V}$ is such that
${U}{V}-\one$ is ${\Tt}$-traceclass and the index satisfies
$$
\Wind({U}\,{V})\;=\;\Wind({U})\,+\,\Wind({V})
\;.
$$
\end{lemma}

\noindent {\bf Proof.} The first claim follows from ${U}{V}-\one=({U}-\one){V}+({V}-\one)$ and the ideal property of the ${\Tt}$-traceclass operators w.r.t. products with covariant operators (note that alternatively, one can use that $\Dd$ is also an ideal). Now by the Leibniz rule and cyclicity 
\begin{eqnarray*}
\Wind({U}\,{V})
& = &
\imath \;\widehat{\Tt}\bigl( ({U}\,{V}-\one)^*(\nabla_1 {U}\,{V}+{U}\,\nabla_1{V})\bigr)
\\
& = &
\Wind({U})\,+\,\Wind({V})
\,+\,
\imath \;\widehat{\Tt}\bigl( (\one-{V})\nabla_1 {U}
\;+\;
(\one-{U})\nabla_1 {V}
\bigr)
\;.
\end{eqnarray*}
But an integration by parts shows that the sum of the last two summands vanishes.
\hfill $\Box$

\section{Proof of Theorem~\ref{theo-bulkinterface}}

Let us introduce a family of Hamiltonians generalizing \eqref{eq-Hamform}:
$$
{H}_{\omega}(\mu)
\;=\;
\widehat{H}_{+,\omega}
\oplus
\widehat{H}_{-,\omega}
\;+\;
\mu\,K_{\omega}
\;,
\qquad
\mu\in[0,1]
\;.
$$
For each of these operators Theorem~\ref{theo-interfacchannels} holds. Hence let $U_\omega(\mu)=\exp(2\pi\imath\,G(H_\omega(\mu)))$. Then $\Pi_{1}U_\omega(\mu)(\Pi_{1})^*$ is a Fredholm operator with $\PP$-almost surely constant index $\Ind(\mu)$. As $U_\omega(\mu)$ is a smooth function of ${H}_{\omega}(\mu)$, it depends continuously on $\mu$ in the norm topology. Hence also the index $\Ind(\mu)$ is constant by the homotopy invariance of the index. Now for $\mu=0$ the Hamiltonian decomposes into a direct sum w.r.t. $\ell^2(\ZM^2)=\Hh_+\oplus\Hh_-$ where $\Hh_\pm=\ell^2(\ZM\times\ZM_\pm)$. Therefore also $U_\omega(0)=\widehat{U}_{+,\omega}\oplus \widehat{U}_{-,\omega}$ with unitary operators $\widehat{U}_{\pm,\omega}=\exp(2\pi\imath\,G(\widehat{H}_{\pm,\omega}))$ on $\Hh_\pm$.

\vspace{.1cm}

Let us now introduce the surjective partial isometries $\Pi_{1,\pm}:\Hh_\pm=\ell^2(\ZM_+\times\ZM) \to \ell^2(\ZM_+\times\ZM_\pm)$. Then $\Pi_1=\Pi_{1,+}\oplus\Pi_{1,-}$ and
$$
\Pi_{1}U_\omega(0)\Pi_{1}
\;=\;
\Pi_{1,+}\widehat{U}_{+,\omega}(\Pi_{1,+})^*\oplus \Pi_{1,-}\widehat{U}_{-,\omega}(\Pi_{1,-})^*
\;.
$$
As both summands on the r.h.s. are Fredholm operators and the index of a direct sum of Fredholm operators is equal to the sum of the indices, it follows that
$$
\Ind(0)
\;=\;
\Ind\big(\Pi_{1,+}\widehat{U}_{+,\omega}(\Pi_{1,+})^*\big)
\;+\;
\Ind\big( \Pi_{1,-}\widehat{U}_{-,\omega}(\Pi_{1,-})^*\big)
\;.
$$
Now both indices correspond to half-space problems and these indices were studied in \cite{KRS} (as well as in \cite{EG}) and shown to be equal to $\Che(P_+)$ and $-\,\Che(P_-)$. This is called the bulk-edge correspondence.  Here the minus sign in $-\,\Che(P_-)$ results from the fact that the operators in the second summand are on the lower half-plane for which the boundary has the opposite orientation. Resuming, the proof of Theorem~\ref{theo-bulkinterface} is concluded by 
$$
2\pi\,\widehat{\Tt}(g(H)J_{1})
\;=\;
\Ind(1)
\;=\;
\Ind(0)
\;=\;
\Che(P_+)
\,-\,
\Che(P_-)
\;.
$$


\end{document}